# Apex-angle-dependent resonances in triangular split ring resonators.

Max A. Burnett[1*], Michael A. Fiddy[1]

[1]Department of Physics and Optical Science, University of North Carolina at Charlotte, North Carolina, USA

*corresponding author, E-mail: mburne10@uncc.edu

## Abstract

Along with other frequency selective structures [1] (circles and squares), triangular split-ring resonators (TSRRs) only allow frequencies near the center resonant frequency to propagate. Further, TSRRs are attractive due to their small surface area [2], comparatively, and large quality factors ($Q$) factors as previously investigated by Gay-Balmaz, et al. [3]. In this work we examine the effects of varying the apex angle on the resonant frequency, the $Q$ factor, and the phase shift imparted by the TSRR element within the GHz frequency regime.

## 1. Introduction

With the advent of the split ring resonator [1], and its unique electromagnetic properties, much work has been done utilizing these structures as band stop or pass elements. These elements have unique resonances depending on their overall shape and material from which they are constructed [5]. An example of a TSRR is shown in Fig. 1.

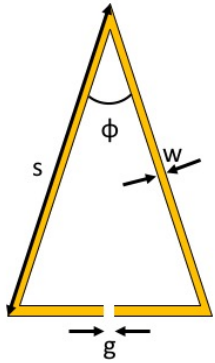

Figure 1. Schematic of a raised element TSRR. The TSRR element is made of copper and backed by FR-4, with parameters s, g, and $\phi$.

Elements fabricated from metal and placed upon a substrate have the ability to manipulate the transmission by acting as band stops through absorption [6]. By etching these elements into a conductive metal sheet with a backing dielectric substrate, a band pass device can be obtained. Not only can these structures be made into useful filters and stops but by manipulating the phase of the incoming radiation, an array of these structures has the ability to act as a flat lens [4]. In this work we analyze the effects of varying the apex angle and gap of TSRRs on the resonance, phase, and Q factor of the resonators. While keeping other dimensions constant (width, $w$, and side lengths, $s$), varying the apex angle and the width of the gap of the TSRR structure causes a shift in the resonance of the structure. Further, we investigate the impact that the width of the gap has on the phase of the radiation as the apex angle's effect has been investigated [4].

### 1.1. Simulation setup

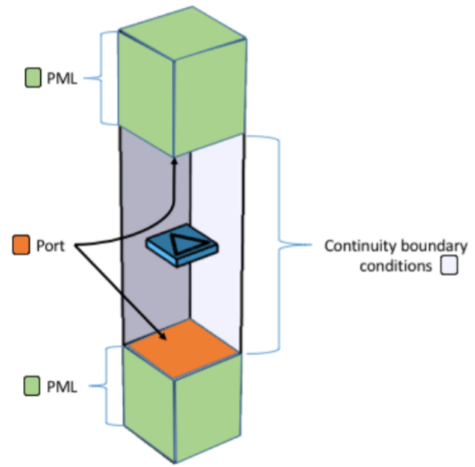

Figure 2. Simulation region. The simulation region is an air cavity backed on all sides by continuous boundary conditions. Residing in the middle of the region is the element and substrate.

The simulation region is depicted within Fig. 2, where a raised element is backed by a dielectric substrate. Copper was chosen as the metal (modeled as a PEC within simulations) for both the raised elements (Fig. 1) and the etched sheet, both 0.025mm thick with varying apex angles, $\phi$, and gap widths, $g$, ($s$ = 7.477mm, $w$ = 0.2mm). While FR-4 was selected as the dielectric substrate with thickness $t$ and side lengths $a_x$ and $a_y$. ($\varepsilon_r$ = 3.85, $t$ = 1.6mm, $a_x = a_y$ = 8.8mm). Both the copper element and substrate slab are set in air and were analyzed using finite element analysis (FEA) through the program COMSOL [6]. The simulation region itself was constructed with sides equal to 15mm and a maximum height of 80mm, with the element and substrate lying in the $xy$-plane, placed in the middle of simulation region. Ports residing on the interior boundaries of the simulation region were used to generate and measure an electric field (6-8 GHz) polarized in $x$-plane while propagating along the $z$-axis. To obtain a better understanding of the TSRR's unique effects, a single raised element is surrounded by continuous boundary conditions to emulate a semi-infinite vacuum around the element and slab. Continuous boundary conditions allow the scattered radiation to exit the simulation region in order prevent self-interaction from reflected radiation. In order to examine the consequences of potential coupling among elements, each side of the unit cell was shortened to sides of $\lambda_R/4$ where $\lambda_R$ is the wavelength corresponding to the resonant frequency of an element. Further, the boundary conditions were changed from continuous to Floquet periodic boundary conditions.

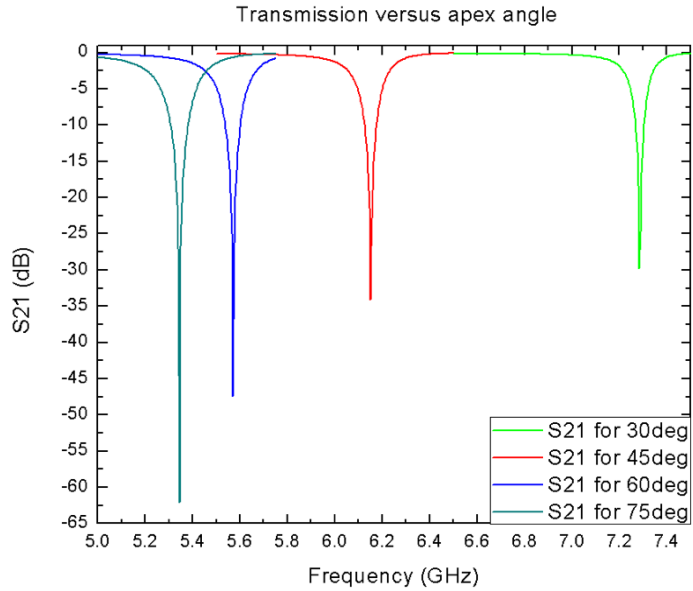

Figure 3. Transmission versus apex angle. The transmission spectra due to varying apex angles of a raised element.

## 2. Results and discussion

Initially, we sought to observe what effects varying the apex angle of the TSRR would have on the transmission through the element and substrate so that we would be able to better understand the consequences of varying $g$. While holding all other parameters constant ($s$, $g$, $w$), we swept through apex angles from 75° to 30° in steps of 15°.

$$f_R \propto \frac{1}{\sqrt{LC}} \qquad (1)$$

In Eq. 1 [4], the resonant frequency ($f_R$) is inversely proportional to the inductance ($L$) and capacitance ($C$) of the element. As indicated in Fig. 3, varying the apex angle of the TSRR results in a nonlinear shift in the resonant frequency of the element. This can be attributed to the inductance of the element changing within Eq. 1. As the apex angle decreases, while holding others fixed, the size of the structure is decreased resulting in a decrease of the inductance. The decrease in the inductance leads to a blue-shift of the resonant frequency. The Q factor was calculated for at each step by:

$$Q = \frac{f_R}{\Delta f}, \qquad (2)$$

where $\Delta f$ is the bandwidth of the element calculated at half power relative to the depth of the resonance.

Table 1. Calculated Q factors for selected apex angles of TSRR

| Apex angle (deg) | Q factor |
|---|---|
| 75° | 82.23 |
| 60° | 101.27 |
| 45° | 124.24 |
| 30° | 224.24 |

The Q factors from the sweep are displayed in Table 1. It is evident that as the apex angle decreases, the Q factor increases as the element nears a more dipole profile.

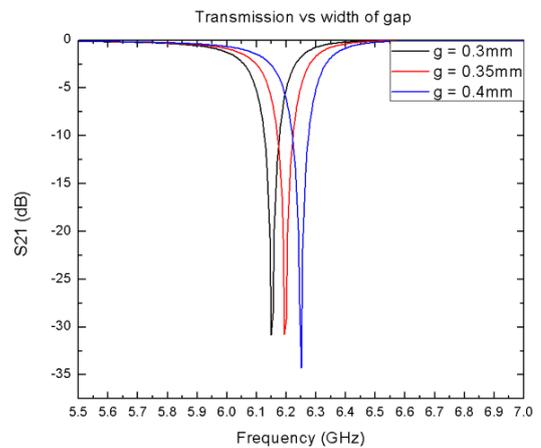

Figure 4. Transmission versus width of gap. The transmission and resonance of a TSRR due to the gap.

After observing the effects of varying the apex angle, we then looked at how the width of the gap, $g$, affected the system; specifically, the resonance and any phase discontinuity the element may impart on the transmitted radiation. This was done so that we could potentially use an array of elements as a flat lens by controlling the phase shift imparted to the transmitted light. As illustrated in Fig. 4, changing the width of the gap causes the resonance of the element to shift. This is due to the fact that the gap introduces a capacitance to the system that varies inversely with width of the gap, from a parallel plate system:

$$C = \epsilon \kappa \frac{A}{w}, \quad (3)$$

where $C$ is the capacitance of the element, $\epsilon$ is the permittivity of the medium in the gap, $\kappa$ is the dielectric strength of the medium, $A$ is the area of the plates, and $w$ is the width or distance between the two plates. By substitution of Equation 3 into Equation 1 we get,

$$f_R \propto \frac{1}{\sqrt{LC}} \propto \sqrt{\frac{w}{L\epsilon \kappa A}}, \quad (4)$$

therein showing that as the width of the gap increases, the capacitance decreases, causing the resonant frequency to increase.

Beyond how the gap alters the resonance, we were interested in its effects, if any, on the phase shift produced by the element. Near resonance, the element will impart a phase shift as shown in Fig. 5.

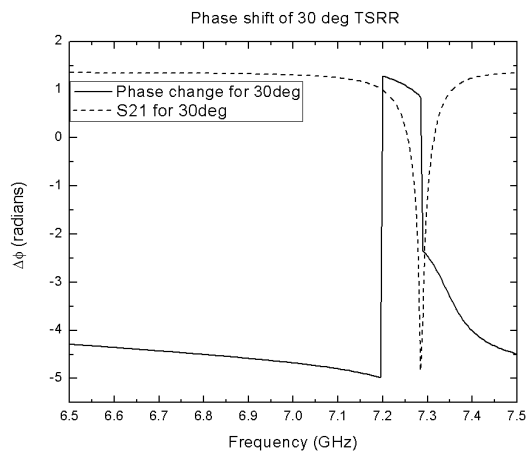

Figure 5. Phase shift of a 30° TSRR. The phase shift is graphed for a 30° TSRR with a gap of 0.3mm. The transmission is superimposed on the graph to show that the phase takes place due to resonance.

The large discontinuity on the left leading edge is simply a phase wrap, but the discontinuity on the right edge is a true phase discontinuity of approximately $\Delta\phi \sim -3.19$ radians.

We then ran a parameter sweep over three values of g: 0.3mm, 1.0mm, and 2.5mm. The results are detailed in Fig. 6.

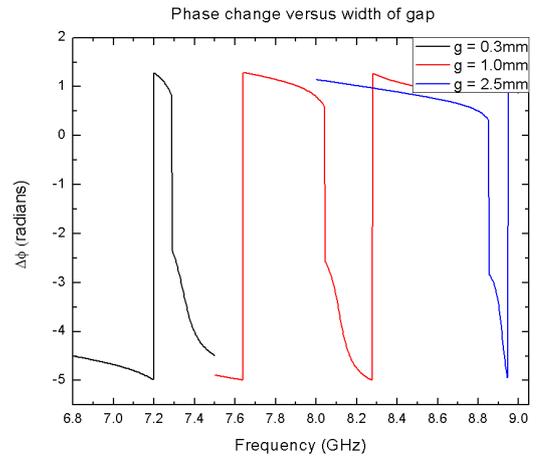

Figure 6. Phase change versus width of gap. The phase shift imparted by the element is simulated for different values of the gap.

The phase shift imparted by the element only changes minimally over the range of gap values, on the order of $1 \times 10^{-2}$. Thus only the resonant frequency and not the phase shift is altered by changing the gap width substantially.

Lastly, we wished to observe the effects of any coupling amongst the elements if the elements were brought closer together. In order to do this, the simulation region was shortened to a length of $\lambda_R/4$ where $\lambda_R$ is the resonant wavelength. Further to allow for full periodicity, the continuous boundaries were replaced with periodic Floquet boundaries to allow the elements to couple.

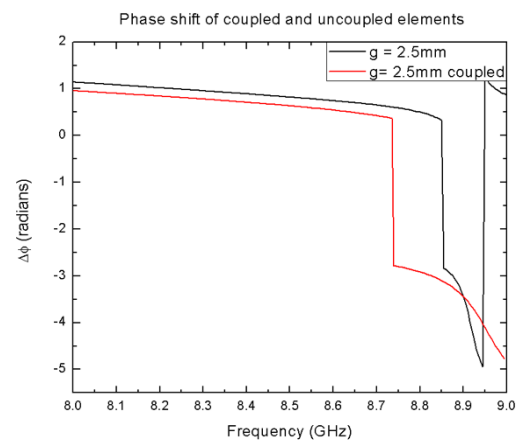

Figure 7. Phase shift of coupled and uncoupled elements. The figure shows the shift in resonance due to the coupling of nearby elements.

Fig. 7 shows that by allowing the elements to couple, the resonant frequency is red-shifted due to coupling but the phase change remains constant over the elements. This is due to the mutual magnetic inductance of the elements as shown in Ref. [3].

## 3. Conclusions

In this paper, we have investigated and shown the effects that the gap of a TSRR has on its resonant frequency and the phase shift that it imparts on the transmitted radiation. It has been shown that the gap itself has the ability to shift the resonant frequency of the element while maintaining a constant phase shift, it is then possible that these elements could be used in a flat lens system. By manipulating the apex angle to impart a specific phase shift as shown in Ref. [4], the gap of a TSRR can then be altered to maintain a specific resonant frequency, thus allowing for another tunable parameter.

## Acknowledgements

The authors would like to acknowledge Dr. Kenneth W. Allen for his helpful insight interpreting the data. Further the authors are grateful for the support from the NSF Center for Metamaterials Award #1068050.